\documentclass[final,3p,times,twocolumn,longtitle]{elsarticle} 

\usepackage{epsfig}

\usepackage{amssymb}

\usepackage[nodots]{numcompress}

\usepackage{lineno}  

\journal{Nuclear Instruments and Methods in Physics Research A}

\begin{document}

\begin{frontmatter}




\title{An improved method for measuring muon energy using the truncated mean of dE/dx}




\author[MadisonPAC]{R.~Abbasi}
\author[Gent]{Y.~Abdou}
\author[Zeuthen]{M.~Ackermann}
\author[Christchurch]{J.~Adams}
\author[Geneva]{J.~A.~Aguilar}
\author[MadisonPAC]{M.~Ahlers}
\author[Berlin]{D.~Altmann}
\author[MadisonPAC]{K.~Andeen}
\author[MadisonPAC]{J.~Auffenberg}
\author[Bartol]{X.~Bai\fnref{SouthDakota}}

% ...
\address[Berlin]{Institut f\"ur Physik, Humboldt-Universit\"at zu Berlin, D-12489 Berlin, Germany}
% ...
\address[Christchurch]{Dept.~of Physics and Astronomy, University of Canterbury, Private Bag 4800, Christchurch, New Zealand}
% ...
\address[Geneva]{D\'epartement de physique nucl\'eaire et corpusculaire, Universit\'e de Gen\`eve, CH-1211 Gen\`eve, Switzerland}
\address[Gent]{Dept.~of Physics and Astronomy, University of Gent, B-9000 Gent, Belgium}
% ...
\address[MadisonPAC]{Dept.~of Physics and Wisconsin IceCube Particle Astrophysics Center, University of Wisconsin, Madison, WI 53706, USA}
% ...
\address[Bartol]{Bartol Research Institute and Department of Physics and Astronomy, University of Delaware, Newark, DE 19716, USA}
% ...
\address[Zeuthen]{DESY, D-15735 Zeuthen, Germany}

\fntext[SouthDakota]{Physics Department, South Dakota School of Mines and Technology, Rapid City, SD 57701, USA}
\fntext[LosAlamos]{Los Alamos National Laboratory, Los Alamos, NM 87545, USA}
\fntext[Bari]{also Sezione INFN, Dipartimento di Fisica, I-70126, Bari, Italy}
\fntext[Goddard]{NASA Goddard Space Flight Center, Greenbelt, MD 20771, USA}
\fntext[corresponding1]{Corresponding author:  Sandra Miarecki, Lawrence Berkeley National Laboratory, 1 Cyclotron Rd, Mail Stop 50R5008, Berkeley CA 94720 USA, Phone +1-510-486-7620, Fax +1-510-486-6738, Email miarecki@berkeley.edu } 
\fntext[corresponding2]{Corresponding author:  Spencer Klein, Lawrence Berkeley National Laboratory, 1 Cyclotron Rd, Mail Stop 50R5008, Berkeley CA 94720 USA, Phone +1-510-486-5470, Fax +1-510-486-6738, Email srklein@lbl.gov }

\begin{abstract}

The measurement of muon energy is critical for many analyses in large Cherenkov detectors, particularly those that involve separating extraterrestrial neutrinos from the atmospheric neutrino background.  Muon energy has traditionally been determined by measuring the specific energy loss ($dE/dx$) along the muon's path and relating the $dE/dx$ to the muon energy.  Because high-energy muons ($E_{\mu}~ >$~1~TeV) lose energy randomly, the spread in $dE/dx$ values is quite large, leading to a typical energy resolution of 0.29 in $\log_{10}(E_{\mu})$ for a muon observed over a 1~km path length in the IceCube detector.  In this paper, we present an improved method that uses a truncated mean and other techniques to determine the muon energy.  The muon track is divided into separate segments with individual $dE/dx$ values.  The elimination of segments with the highest $dE/dx$ results in an overall $dE/dx$ that is more closely correlated to the muon energy. This method results in an energy resolution of 0.22 in $\log_{10}(E_{\mu})$, which gives a 26\% improvement.   This technique is applicable to any large water or ice detector and potentially to large scintillator or liquid argon detectors. 

\end{abstract}


\end{frontmatter}
%

\section{Introduction}

Large ice or water Cherenkov detectors may observe up to 100,000 $\nu_{\mu}$ per year \cite{Halzen:2010yj}.  These events are used for a wide variety of analyses, including searches for point sources of neutrinos \cite{IceCube:2011ai,Abbasi:2010rd, Bogazzi:2011zza}, diffuse extraterrestrial neutrinos \cite{Bogazzi:2011zza,Abbasi:2011jx,Dzhilkibaev:2009ja}, standard and non-standard neutrino oscillations \cite{Abbasi:2010kx}, and measurements of the total neutrino-nucleon cross-section via neutrino absorption in the Earth.

These analyses rely upon the measurement of the neutrino energy, which is determined from the energy of the muon that is created in the neutrino interaction.  Above $\sim$1~TeV, the muon energy is usually determined by measuring the specific energy loss, $dE/dx$, of the muon as it travels through the detector.  The Cherenkov photons from the muon and also those derived from the charged particles produced by stochastic (random) muon interactions are then detected.  This approach is disadvantageous because, for $E_{\mu}>1$~TeV, muons lose most of their energy stochastically, and a small number of high-energy interactions will not only skew the mean but also enlarge the spread in $dE/dx$ values. 

In this paper, we present an improved method for calculating the muon energy loss, which leads to significant improvement in the energy resolution.  Instead of averaging the muon $dE/dx$ over the entire observed muon path length, we divide the path into independent segments, or bins.  The $dE/dx$ is calculated separately for each bin.  Then, the bins with the highest $dE/dx$ values are discarded before calculating a new average $dE/dx$, thus producing a truncated mean.  This method is successful because the truncated mean minimizes the effects of the large stochastic events which would otherwise skew the mean and enlarge the spread.  

This is the first time that the truncated mean has been systematically applied to the energy measurement of high-energy muons.  The truncated mean method has previously been calculated for muons \cite{Auchincloss:1993zu}, although that analysis did not use it to determine the muon energy.  The method was also explored at a basic level for the DUMAND project \cite{Mitsui:1992nt}.  The method has parallels to the one that was used to identify pions, kaons, and protons in wire chambers from their specific energy loss $dE/dx$ in the gas.  By discarding the highest 30\% or 40\% of the $dE/dx$ measurements from the wires, the energy resolution was greatly improved \cite{Hauschild:1996gv,Bichsel:2006cs}.  As with wire chamber $dE/dx$ measurements, the muon energy resolution is improved by discarding the most energetic stochastic losses.  This method should be applicable to any large water or ice Cherenkov detector, such as IceCube \cite{Halzen:2010yj}, ANTARES \cite{Collaboration:2011nsa}, or the proposed KM3NeT \cite{Sapienza:2011zzb}, MEMPHYS \cite{Agostino:2012fd}, or Hyper-Kamiokande (Hyper-K) \cite{Abe:2011ts}.  This approach may also be useful for proposed scintillator or liquid argon detectors, such as the Low Energy Neutrino Astronomy detector (LENA) \cite{Wurm:2011zn} or the Long-Baseline Neutrino Experiment (LBNE) \cite{Barker:2012nb}.  

\section{Muon Energy Loss}

Muons lose energy via ionization and by stochastic processes such as bremsstrahlung, pair production, and photonuclear interactions \cite{Sakumoto:1991xn,Nakamura:2010zzi,mmc,Groom:2001kq}.  Muons also emit Cherenkov radiation, but this is a very small fraction of the total energy loss.  Ionization loss is roughly independent of muon energy and essentially constant per unit length, while the $dE/dx$ due to stochastic processes rises linearly with the energy.  Thus, the total average energy loss of the muon is \cite{Lipari:1991ut}
\begin{equation}
\frac{dE_{\mu}}{dx} = A + BE_{\mu}
\label{eq:dEdx}
\end{equation}
where $A \approx$ 0.0024~GeV {(g/{cm}$^2$)}$^{-1}$ accounts for the energy loss due to ionization, and $B \approx$ 0.000032 {(g/{cm}$^2$)}$^{-1}$ is due to the stochastic energy loss \cite{mmc}.  Ionization effects are dominant for muons below $\sim$500~GeV, at which point the sum of bremsstrahlung, pair production, and photonuclear effects is larger than ionization.  As muon energy increases to $\sim$10 TeV, ionization effects account for only $\sim$9\% of the total muon energy loss, with a much reduced contribution as muon energy increases \cite{mmc}.  

Two main methods are used to measure $E_{\mu}$.  For lower-energy muons, the energy is proportional to the path length. This approach works if the muon is contained within the detector, and the starting and stopping points can be determined.  For example, a 300~GeV muon has a most probable path length of $\sim$1~km in ice. The most probable path length is $\sim$2~km for a 1~TeV muon, and the length rises only logarithmically with $E_{\mu}$, reaching 20~km for a 1~PeV muon \cite{mmc}. 

At energies above a few hundred GeV, one measures the $dE/dx$ and infers $E_{\mu}$ from that value, either by using Eq.~\ref{eq:dEdx} as an approximation or by calculating the correlation using Monte Carlo simulation.  The total energy loss in the entire detector is summed, either using calorimetric measurements of $dE/dx$ (such as by the Frejus collaboration) \cite{Daum:1994bf} or by observing the light produced by the stochastic interactions of the muon and charged particles \cite{Abbasi:2010ie}.   

Muon energy loss is a complex sequence of events where the stochastic processes create many secondary particles that are also subject to further interactions that can result in additional charged particles.  The charged particles produce Cherenkov light with a flux that is proportional to the particle energy.

The IceCube Neutrino Detector is highly suited for a study involving muon energy loss.  This analysis used IceCube in its 2010 configuration, when the detector consisted of 79~vertical strings (with roughly 125~m horizontal spacing in a triangular grid), with 60~optical modules per string (with 17~m vertical spacing) between 1450 and 2450 m below the surface of the ice sheet at the South Pole. This included DeepCore, which is a denser subarray of strings near the detector center with more-sensitive optical modules placed preferentially on the bottom 400~m of each string \cite{Collaboration:2011ym}. Each optical module holds a 25~cm photomultiplier tube (PMT), plus associated electronics for data acquisition, control, and calibration \cite{Halzen:2010yj}.

Simulated muon trajectories were determined using a program called neutrino-generator (NUGEN) which is based upon the All Neutrino Interaction Simulation (ANIS) \cite{Gazizov:2004va}. NUGEN simulation includes several effects, including the ice/rock boundary below the detector, Earth-neutrino absorption using the density profile from the Preliminary Reference Earth Model (PREM) \cite{Dziewonski:1981xy}, neutral current regeneration, etc., but it does not include neutrino oscillations.  The program starts with an isotropic distribution of neutrinos on the Earth's surface, following, in this case, a $E_{\nu}^{-1}$ spectrum for 10~GeV~$<E_{\nu}<$~1~EeV.  These neutrino events can be reweighted to simulate softer spectra.  NUGEN then propagates the neutrinos to the South Pole, accounting for neutrino absorption in the Earth.  Finally, NUGEN simulates the neutrino interactions.  The resulting muons are propagated through the ice using the Muon Monte Carlo (MMC) code \cite{mmc}, which includes detailed models for muon energy loss and the Landau-Pomeranchuk-Migdal effect \cite{Klein:1998du}.

From the energy deposition determined by MMC, a program called Photonics determines the number and timing distribution of expected photoelectrons for a given optical module \cite{Lundberg:2007mf}.  Photonics also accounts for the detector geometry of the optical modules.   Finally, a set of detector simulation programs models the response of the PMT, electronics, and triggers \cite{Ahrens:2003fg}.  Then the simulated data are analyzed using standard IceCube reconstruction programs.  Muon tracks are typically reconstructed with an angular resolution of~$<$~1~degree \cite{moon shadow} and a positional accuracy (in the plane perpendicular to the track) of a few meters.  

Detailed simulation verification studies at a low-level \cite{Achterberg:2006md} and high-level \cite{Abbasi:2010ie} have been published.  The systematic uncertainty in simulation is largely due to the understanding of the optical properties of the ice, with smaller contributions from the angular acceptance and saturation effects of the optical sensors.  An analysis using in situ light sources \cite{Ackermann} discusses the ice properties in detail, with a new paper from IceCube forthcoming.  In addition, simulations were completed using 3 different ice models that resulted in a shifting of the mean but no degradation in the energy resolution, with an overall maximum variation of 13\% in the energy reconstruction.

\subsection{Conventional $dE/dx$}
\vspace{0.04in}
Before discussing the truncated mean $dE/dx$ in detail, we will discuss the more conventional approach.  First, the number of photoelectrons observed by all of the optical modules is summed.  Then, the expected number of photoelectrons for a fixed energy loss of 1~GeV/m with the same trajectory is calculated.  Since the photoelectron yield is expected to be directly proportional to the muon energy loss rate, the calculated $dE/dx$ value is equal to the ratio of the observed number of photoelectrons to the expected number for the 1~GeV/m track, multiplied by 1~GeV/m.  This linear relationship is a proper $dE/dx$ approximation for muon energies above $\sim$1~TeV.  Because the method uses the distance between each sensor and the muon track, it automatically accounts for varying track geometries, sensor densities, and optical properties of the detector medium, since those effects are included in both the observed and expected numbers of photoelectrons.  

\begin{figure}[!h]
\centering
\includegraphics[width=0.48\textwidth]{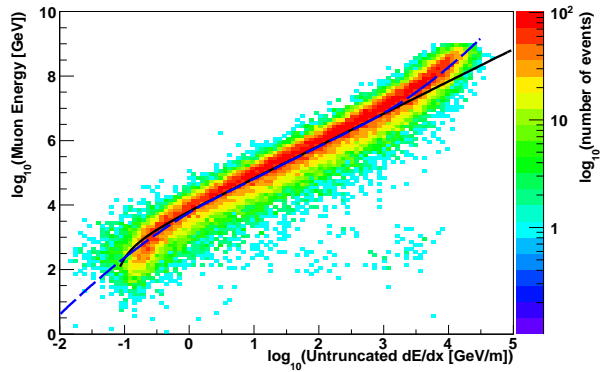} 
\caption{(color online) Simulated $E_{\mu}$ vs. conventional (untruncated) $dE/dx$ at the center of the detector.  The $dE/dx$ value is calculated from the ratio of the observed energy deposition to the expected deposition for a minimum ionizing muon with the same trajectory.  The solid line uses Eq.~\ref{eq:dEdx} for the ``linear'' fit.  The dashed line uses the 3-equation fit described by Eq.~\ref{eq:log} and Eq.~\ref{eq:linear}. }
\label{fig:log_orig}
\end{figure}

Figure~\ref{fig:log_orig} shows a scatter plot of simulated muon energy versus calculated $dE/dx$.  Because the entire track is used, the simulated muon energy value comes from the track's closest approach to the center of the detector (instead of at the entrance to or exit from the detector volume).  The $dE/dx$ values have a large spread. Some of the spread is due to the limited sampling by the sensors and consequently the uncertain measurement of the energy loss.  However, most of the spread is from the event-by-event variation in stochastic energy losses.  
\begin{figure}[!h]
\centering
\includegraphics[width=0.50\textwidth]{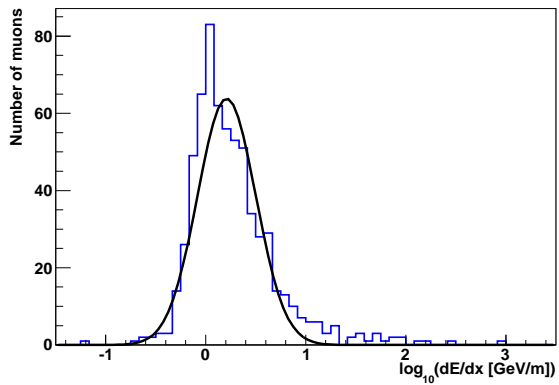} 
\caption{(color online) Conventional (untruncated) muon $dE/dx$, for muons with simulated energies $\sim$10~TeV ($\log_{10}(E_{\mu}$) = 4.0 $\pm$ 0.05).  The black line is a Gaussian fit, with $\sigma$ = 0.29 $\pm$ 0.01, and RMS = 0.44.  The excess of events to the right of the Gaussian curve are muons with large stochastic losses. }
\label{fig:orig_dedx}
\end{figure}

The spread in $dE/dx$ values for a given true energy is quite large for the conventional (untruncated) method.  Figure~\ref{fig:orig_dedx} shows a typical example of the spread for a narrow energy slice, with this one at $\sim$10~TeV.  The excess of events to the right of the Gaussian curve is a direct result of the large stochastic losses.

The desired output of the method is the energy for each muon. This is done using a fit to the curve in Fig.~\ref{fig:log_orig} to calculate $E_{\mu}$ from the $dE/dx$ value, and then we can compare the calculated energy to the simulated energy.  

It is apparent that the ``linear'' fit curve from Eq.~\ref{eq:dEdx} does not follow the data precisely, particularly at higher energies.  The main experimental effect is saturation in the PMT at very high light intensities \cite{Abbasi:2010vc}.  During the reconstruction of both actual and simulated data, no correction is applied for PMT saturation.  This produces the upturn in the curve at higher energies for $dE/dx$ ${\scriptstyle \gtrsim}$ $10^{3}$~GeV/m.  For $dE/dx$ ${\scriptstyle \lesssim}$ 1~GeV/m, there is also a small downturn in the curve, making it S-shaped.  This occurs because the muon range at lower energy is less than the size of the detector, while the truncated method assumes that the track is infinite. 

We examined a variety of simple fit equations to improve our ability to characterize the muon energy resolution.  The optimal fit had three separate sections:  a second-order logarithmic ($\log_{10}$) polynomial fitted to the lower energy curve (for $dE/dx$~${\scriptstyle \lesssim}$~1.5~GeV/m) and another second-order logarithmic polynomial to the higher energy curve ($dE/dx$~${\scriptstyle \gtrsim}$~500~GeV/m): 
\begin{equation}
\log (E_{\mu}) = A' + B'  \log \left(\frac{dE}{dx}\right) + C' \log^2 \left(\frac{dE}{dx}\right)
\label{eq:log}
\end{equation}
where $E_{\mu}$ is the energy of the muon (in GeV) at the track's closest approach to the center of the detector, $dE/dx$ is the calculated value (in GeV/m), and $A^{\prime}$, $B^{\prime}$, and $C^{\prime}$ are dimensionless constants from the best fits to the curves.  For the mid-range, we used the logarithmic version of the linear $dE/dx$ equation (Eq.~\ref{eq:dEdx}):  
\begin{equation}
\log (E_{\mu}) = \log \left(\frac{1}{B''} \left[ \frac{dE}{dx} - A'' \right]  \right)
\label{eq:linear}
\end{equation} 
where $E_{\mu}$ is the energy of the muon (in GeV), $dE/dx$ is the calculated value (in GeV/m), and $A^{\prime\prime}$ and $B^{\prime\prime}$ are dimensionless constants from the best fit to the curve.  Because of PMT saturation and other detector effects, the constants in Eq.~\ref{eq:linear} must be adjusted for a proper fit.  The new fit curve is shown in Fig.~\ref{fig:log_orig} as a dashed line.  We used this 3-equation fit throughout this analysis, instead of the linear fit.

\begin{figure}[!h]
\centering
\includegraphics[width=0.50\textwidth]{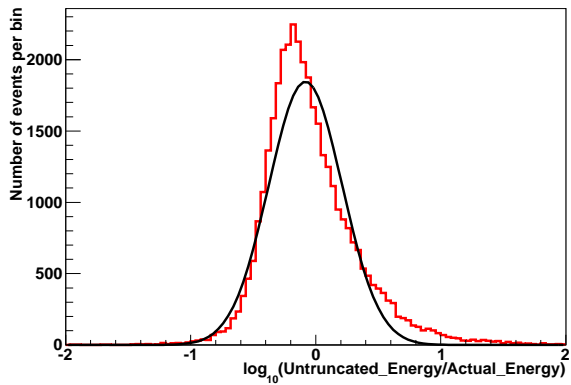} 
\caption{(color online) Conventional (untruncated) muon energy resolution, using the 3-equation fit curve from Fig.~\ref{fig:log_orig}, for muons with simulated energies between 1~TeV and 1~EeV.  The black line is a Gaussian fit, with the peak offset by -0.08 in $\log_{10}(E_{\mu})$, $\sigma$ = 0.29, and with an RMS of 0.37.  The excess of events to the right of the Gaussian curve is a direct result of the large stochastic losses. }
\label{fig:orig_res}
\end{figure}

The energy residual $E_{res}$ is nominally defined as $E_{reco}~-~E_{sim}$, where $E_{reco}$ and $E_{sim}$ are the reconstructed and actual simulated energies, respectively, at the track's closest approach to the center of the detector.  Since this analysis covered a wide range of energies, the energy residual was redefined as $E_{res}~=~\log_{10}(E_{reco})~-~\log_{10}(E_{sim})$.  Figure~\ref{fig:orig_res} shows the results of this energy comparison, using the 3-equation fit.  The distribution is skewed, with a long tail of events containing large stochastic energy deposition, indicating that the energies for these events are badly overestimated.  About 5.4\% of the muon energies are overestimated by more than a factor of 5.  

If the distribution of energies is fit to a Gaussian, the energy resolution $\sigma$ is 0.29 in $\log_{10}(E_{\mu})$, or roughly a factor of 2.0 in energy.  The statistical error is less than 1\% for this resolution and all other resolutions presented in this paper (unless otherwise noted). However, the peak of the curve is offset from 0 by -0.08, and the curve is skewed.  The shift towards lower muon energies occurs because the fit curve in Fig.~\ref{fig:log_orig} does not line up with the highest concentration of events. This occurs because the fit curve is pulled towards higher $dE/dx$ values for a given muon energy due to the stochastic losses.  For a specific $dE/dx$ value, the calculated energy using the fit equations for the bulk of the events will be too low.  This pushes the peak of the distribution of energy residuals toward the negative.  Fortunately, in a full analysis, this bias would cancel out if the method were used identically for both data and simulation, and the spread would be the most relevant issue.   

In this analysis, we used both the Gaussian $\sigma$ and the RMS to characterize the improvements in the spread in calculated $dE/dx$ and $E_{\mu}$.  In most cases, the RMS was significantly larger than $\sigma$ because the RMS is more sensitive to the high-energy tails in the distributions.  

\subsection{Truncated Mean $dE/dx$}
\vspace{0.04in}
The truncated mean method divides the track into bins which are bordered by planes perpendicular to the track, and the sensor data are binned accordingly by their location.  The $dE/dx$ value is determined as before by finding the ratio of the observed photoelectrons to the expected photoelectrons, but a separate ratio is determined for each bin instead of the event as a whole.  Then a fraction of the bins (using the optimization discussed below) with the highest ratios are discarded, and the truncated $dE/dx$ is calculated by summing the remaining observed photoelectrons and expected photoelectrons and creating a new ratio. 

Many factors influence the choice of bin size, with the objective to create bins as independent from each other as possible. Large Cherenkov detectors use sparse sampling with sizeable distances between sensors, and each sensor observes light over a limited track segment. The length of the segment depends on the optical properties of the medium, with different properties for sea water, ice, and liquid scintillators that vary with the wavelength of light.  

In water, absorption effects are much stronger than scattering.  Thus the most appropriate measure of optical properties would be the absorption length, which peaks at wavelengths $\sim$470~nm.  The absorption length is $\sim$22~m in lake water \cite{Balkanov:1999uq} and $\sim$60~m in seawater \cite{Aguilar:2004nw}.  

In deep glacial ice, however, scattering effects are stronger than absorption, and the appropriate measure is the light propagation length, which peaks $\sim$400~nm.  This length depends on both the effective scattering length $\lambda_{es}$ and the absorption length $\lambda_{a}$ in the medium, using the relation
\begin{equation}
\lambda_{prop} = \sqrt{ \frac{\lambda_{es} \lambda_{a} } {3} } 
\label{eq:propagation}
\end{equation} 
In IceCube, for example, the average absorption coefficient is $\sim$0.008~m$^{-1}$, or an absorption length of 125~m, while the average effective scattering coefficient is $\sim$0.040~m$^{-1}$, or an effective scattering length of 25~m \cite{Ackermann}.  The average propagation length is thus $\sim$32~m, which increases to $\sim$45~m for the very clear ice in the bottom of the detector.  

For liquid scintillator detectors, the absorption length is the appropriate measure of optical properties, which is typically 10-20 m \cite{Wurm:2011zn}.

The detector geometry may also impose constraints on the truncated mean's binning method.  In ANTARES, for example, the string spacing is 70~m, which is more than the reported absorption length.  In Hyper-K, with its large photocathode coverage, the water properties may be the determining factor for the optimal bin length.  Hyper-K's 250-m tank length is far longer than the light attenuation length, and the detector geometry will naturally allow for track segmentation.

Most of the vertical strings of optical modules in IceCube are deployed on a 125-m grid, making the detector roughly hexagonal in shape.  For near-horizontal tracks, this string separation sets the minimum sampling distance, and hence a minimum bin length.  In this study, since the IceCube array is not perfectly regular and to avoid having 0 or 2 strings in a bin for horizontal tracks, a 120-m bin length was used, spreading the energy more evenly across the bins. Due to the size of the detector, events could have up to 15 bins.  For comparison, the average distance between the occurrence of any stochastic event, where the muon loses more than 1\% of its energy, is $\sim$250 m at 1~TeV, using the MMC program for ice \cite{mmc}.  This distance is much larger than the bin size.

\begin{figure}[!h]
\centering
\includegraphics[width=0.48\textwidth]{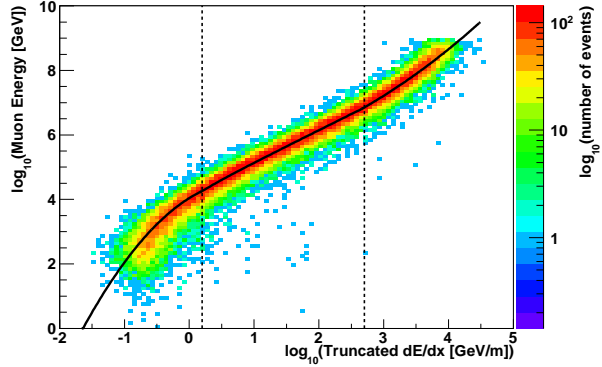} 
\caption{(color online) Simulated $E_{\mu}$ vs. truncated $dE/dx$ at the center of the detector, for the optimized 40\% truncation. The distribution is much narrower than in Fig.~\ref{fig:log_orig}. The vertical dashed lines identify the ranges for the three fit equations (represented by the solid black line).  }
\label{fig:log_bins}
\end{figure}

Figure~\ref{fig:log_bins} shows a scatter plot of simulated $E_\mu$ versus truncated $dE/dx$ for the same events as in Fig.~\ref{fig:log_orig}, but now using the truncated $dE/dx$ with 40\% of the highest bins discarded, and including the optimizations discussed in Section \ref{section:optimize}.  For this analysis, a minimum of 3 bins was required for the event to qualify, since a 40\% truncation would result in 0 bins omitted for a 2-bin event.  The distribution is much narrower than the conventional untruncated $dE/dx$ method but retains the same S-shape, for the same reasons.    

\begin{figure}[!h]
\centering
\includegraphics[width=0.50\textwidth]{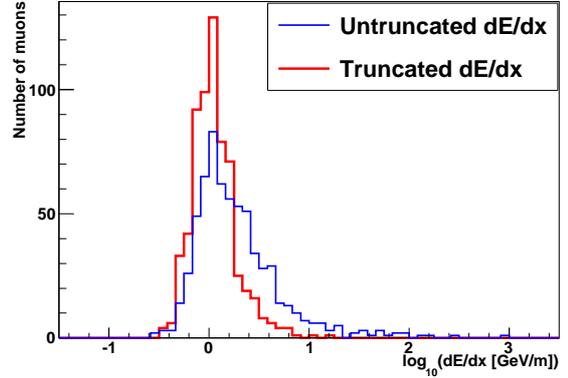} 
\caption{(color online) Truncated muon $dE/dx$ vs. conventional (untruncated) muon $dE/dx$, for simulated energies $\sim$10~TeV ($\log_{10}(E_{\mu}$) = 4.0 $\pm$ 0.05).  For the truncated method, the Gaussian $\sigma$~=~0.18 $\pm$ 0.01, and RMS~=~0.22, compared to the untruncated method values of $\sigma$~=~0.29 and RMS~=~0.44.  The effect of the large stochastic events has been greatly diminished, thus reducing the spread.  }
\label{fig:orig_bins_dedx}
\end{figure}

The spread in $dE/dx$ values is much narrower for the truncated method than for the conventional method.  A comparison of the distributions of $dE/dx$ values for the same narrow energy range as in Fig.~\ref{fig:orig_dedx} is shown in Fig.~\ref{fig:orig_bins_dedx}.  For the truncated method, the Gaussian $\sigma$ is 0.18, compared to the untruncated value of 0.29, which constitutes a significant improvement, and the skewness is much reduced.

\begin{figure}[!h]
\centering
\includegraphics[width=0.50\textwidth]{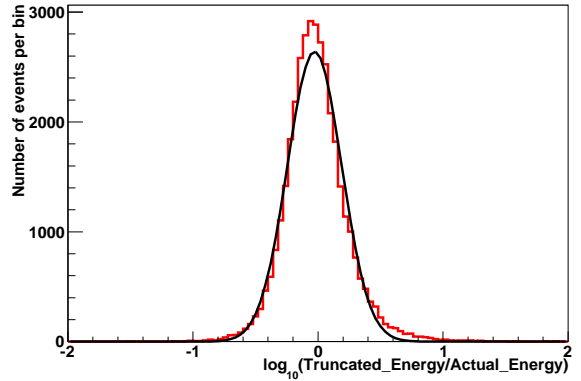} 
\caption{(color online) Truncated muon energy resolution, using the fit curve from Fig.~\ref{fig:log_bins}, for muons with simulated energies between 1~TeV and 1~EeV.  The number of entries to the right of the Gaussian curve (black line) is much smaller with the truncated method, as compared to Fig.~\ref{fig:orig_res}. The energy resolution improved to $\sigma$ = 0.22 in $\log_{10}(E_{\mu})$ (a factor of 1.6 in energy), with the mean of the Gaussian curve slightly offset by -0.03, and an RMS value of 0.25. }
\label{fig:bins_res}
\end{figure}

Figure~\ref{fig:bins_res} shows the corresponding energy resolution using the 3-equation fit from Fig.~\ref{fig:log_bins} to calculate the energy.  Compared to Fig.~\ref{fig:orig_res}, the high-energy tail is much smaller in the truncated plot, with only 1.3\% of the events reconstructed with more than 5 times their actual energy.  The Gaussian resolution has improved to 0.22 in $\log_{10}(E_{\mu})$, which is a factor of 1.6 as compared to 2.0 for the conventional untruncated method (with the 3-equation fit).  

\section{Algorithm Optimization and Performance} \label{section:optimize}

We studied several variations of the algorithm to optimize performance.  The parameters included the fraction of bins discarded in the truncated mean (Section \ref{section:percentage}) and limits on the distance between the sensor and the reconstructed track (Section \ref{section:distance}).  We also evaluated the inclusion or exclusion of the unhit sensors in the calculation (Section \ref{section:hit}), and we explored a variation in the truncated method that treated each sensor as its own bin (the ``DOMs method'') (Section \ref{section:doms}).  Lastly, we looked at the median versus the mean for both the binning and DOMs methods (Section \ref{section:median}).  
 
\subsection{Truncation Percentage} \label{section:percentage}
\begin{figure}[!h]
\centering
\includegraphics[width=0.50\textwidth]{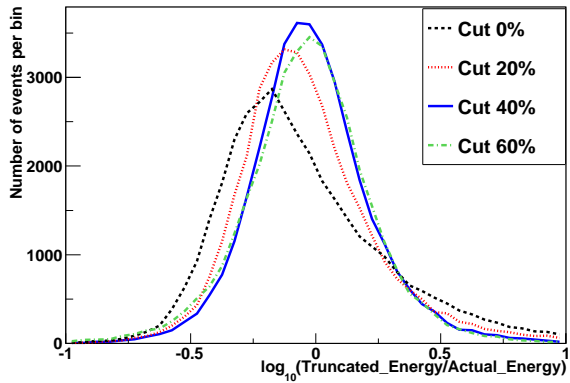} 
\caption{(color online) Energy resolution for truncation percentages ranging from 0\% to 60\%, for $E_{\mu}$~$>$~1~TeV.  The optimal truncation percentage was 40\%.  Gaussian $\sigma$ were 0.27, 0.23, 0.22, and 0.23 for 0, 20, 40, and 60\%, respectively.  The optimization techniques described in Sections \ref{section:distance} and \ref{section:hit} have already been applied. }
\label{fig:percent_truncation}
\end{figure}

\vspace{0.04in}
The optimal truncation was determined by evaluating truncation percentages between 0 and 80\% of the bins. If the percentage of cut bins resulted in a non-integer value, the fraction was rounded down (for example, 40\% of 4 bins would result in 1 bin cut).  Figure~\ref{fig:percent_truncation} shows the resulting energy resolutions for different truncating percentages (up to 60\% shown).  In this analysis, 40\% was the optimal cut.  

\subsection{Distance of optical modules from track} \label{section:distance}
\vspace{0.04in}
Energy measurement by $dE/dx$ relies on an accurate measurement of the distance between the tracks and the optical modules.  However, the reconstructed track positions (transverse to the direction of motion) can be off by several meters.  Also, data from optical modules that are very far from the track are less useful because of the increased chance that the sensor hits are noise.  

To avoid these issues, we evaluated the performance of the method when requiring a minimum and maximum distance between the track and the sensors.   The resolutions for various distances are shown in Table \ref{table:table1} for horizontal muons with simulated energy~$>$~10~TeV.  The optimal energy resolution used optical modules that were located between 10 and 80 m from the track.   In an optically noisy environment, such as seawater, a smaller maximum distance likely would be optimal. 
\begin{table}[!h] \scriptsize
\centering \begin{tabular}{|c||c|c|c|c|c|c|c|}
\hline 
~& \multicolumn{7}{|c|}{ Cylinder distance (meters)} \\
\hline Cuts & None & 0-100 & 10-100 & 0-80 & 10-80 & 0-60 & 10-60 \\
\hline 
  0\% & 0.362 & 0.345 & 0.331 & 0.342 & 0.328 & 0.329 & 0.308 \\
10\% & 0.324 & 0.316 & 0.282 & 0.307 & 0.291 & 0.319 & 0.298 \\
20\% & 0.276 & 0.273 & 0.222 & 0.244 & 0.216 & 0.242 & 0.221 \\
30\% & 0.255 & 0.231 & 0.211 & 0.228 & 0.206 & 0.229 & 0.204 \\
40\% & 0.248 & 0.227 & 0.210 & 0.229 & {\bf0.198} & 0.225 & 0.204 \\
50\% & 0.250 & 0.234 & 0.215 & 0.229 & 0.200 & 0.227 & 0.202 \\ 
\hline \end{tabular}
\caption{Energy resolution $\sigma$ in $\log_{10}(E_{\mu})$ for varying truncated percentage cuts and cylinder distances (in meters) from track, for muons with simulated energy~$>$~10~TeV.  The optimal resolution occurred with 40\% truncation and a cylinder of 10 to 80 m.  When the lower-energy muons are included ($E_{\mu} >$~1~TeV), the energy resolution is slightly degraded ($\sigma$~=~0.22). } 
\label{table:table1} 
\end{table}

\subsection{Optical modules without observed photoelectrons} \label{section:hit}
\vspace{0.04in}
In IceCube, most energy reconstruction methods (such as the truncated mean method) use a Gaussian error model, which is valid at high energy but not at low energy where the number of detected photons is bounded at the low end by zero but is not bounded above. For optical modules that detect many photons, this is mostly irrelevant because Poisson errors become symmetrical. Therefore, the inclusion of many low-amplitude sensors, such as those that did not observe any photons (the ``unhit'' sensors), could bias the likelihood calculation.

In addition, isolated hits (optical modules without hits in their nearest or next-to-nearest neighbors) are likely to be noise.  For these reason, these energy-loss calculations only used the data from optical modules where the nearest or next-to-nearest neighbors were hit within $\pm~1~\mu$s \cite{Abbasi:2008aa}, otherwise known as coincident-hit modules.  

In calculating the number of expected photoelectrons, there is no distinction between hit and unhit optical modules, and Photonics will return non-zero amplitudes for all optical modules.  To reduce the effect of this mismatch,  we studied the effect of counting the expected photoelectrons from only the optical modules with coincident hits, rather than using all of the optical modules.

\begin{figure}[!h]
\centering
\includegraphics[width=0.48\textwidth]{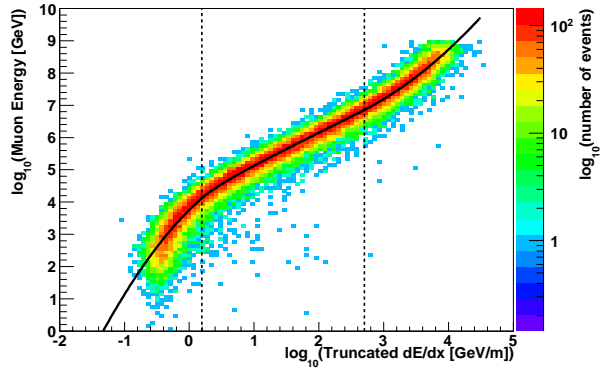} 
\caption{(color online) Simulated $E_{\mu}$ vs. truncated $dE/dx$, using only the ``hit'' optical modules within the cylinder.  The vertical dashed lines identify the ranges for the three fit equations (represented by the solid black line).  At low $dE/dx$ values, the curve drops off more steeply than in Fig.~\ref{fig:log_bins}, which adversely affects the energy resolution in this range.  }
\label{fig:only_hit}
\end{figure}

Figure~\ref{fig:only_hit} shows the results from this alternate methodology combined with the truncated mean.  The energy resolution for this ``hit''  technique was slightly better at the higher energies.  However, at lower energies, the curve is much steeper than in the previous ``all-sensor'' truncated curve, and the energy resolution was significantly worse when using only the hit sensors in the $dE/dx$ calculation.  Thus, this ``hit approach'' was not used in the final truncated method.  Instead, the data from the coincidence-hits (within the cylinder) were used for the actual photoelectron counts, and all the sensors (hit and unhit) in the cylinder were used for the calculation of expected photoelectrons.

\subsection{DOMs method} \label{section:doms}
\vspace{0.04in}
An additional option was explored that diverged from the binning method, which was nicknamed the ``DOMs method''  after the digital optical modules in IceCube.  In this option, instead of using 120~m bins, each optical module became a ``bin'' and was given its own $dE/dx$ value by dividing the observed number of photoelectrons by the expected number of photoelectrons.  The advantages of this method are (1) the track can be shorter (no minimum number of bins), (2) additional data will be used in the energy determination, and (3) the algorithm is slightly faster.  The main disadvantage is that a large stochastic event would be recorded by many sensors, and thus there would be an undesirable correlation between sensors. 

The technique was optimized (using similar procedures to those outlined above) to require a minimum of 8 optical modules within 0 to 60 m of the reconstructed track for the event to qualify (versus 10 to 80 m for the bins method), and the optimal truncation was to omit the highest 50\% of the optical modules (versus 40\% of the bins).  Then the $dE/dx$ values for the remaining optical modules were averaged.  This yielded an energy resolution and RMS improvement that was similar to that of the binning method ($\sigma$~=~0.22 and RMS~=~0.25). 
\begin{figure}[!h]
\centering
\includegraphics[width=0.48\textwidth]{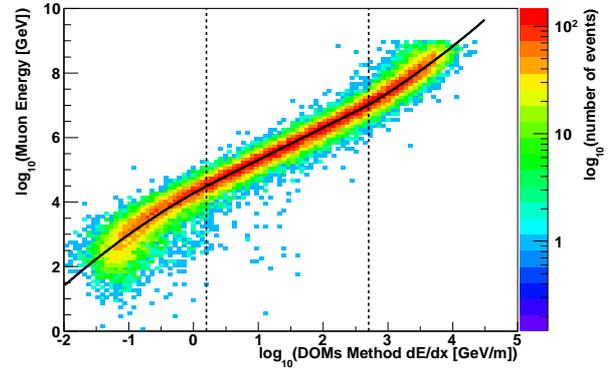} 
\caption{(color online) Simulated $E_{\mu}$ vs. truncated $dE/dx$ (DOMs method) at the center of the detector.  The vertical dashed lines identify the ranges for the three fit equations (represented by the solid black line).  The event distribution is much narrower than in Fig.~\ref{fig:log_orig} but similar to the binning method in Fig.~\ref{fig:log_bins}.  }
\label{fig:log_doms}
\end{figure}
\begin{figure}[!h]
\centering
\includegraphics[width=0.50\textwidth]{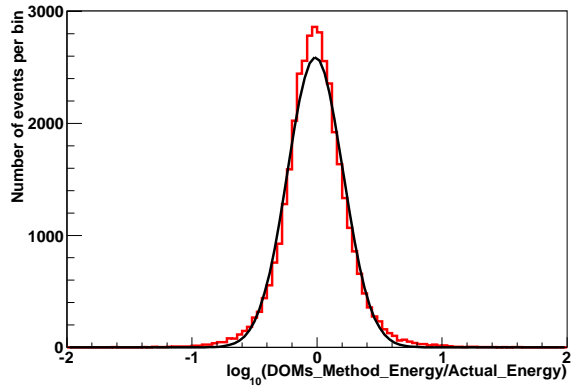} 
\caption{(color online) Truncated muon energy resolution for the DOMs method, for muons with simulated energies between 1~TeV and 1~EeV, using the fit curve from Fig.~\ref{fig:log_doms}.  The number of events to the right of the Gaussian curve is much smaller than in Fig.~\ref{fig:orig_res}. The resolution is very similar to that of the optimized binning method as shown in Fig.~\ref{fig:bins_res} ($\sigma$~=~0.22 and RMS~=~0.25). }
\label{fig:doms_res}
\end{figure}

Figures~\ref{fig:log_doms} and \ref{fig:doms_res} show the results of this method.  One or the other method may prove more useful in the various detector designs and is an option that can be explored.  In this analysis, about 4\% of the events would qualify for only one of the methods, while the remainder would qualify for both.  Including the two approaches maximizes the capability of the truncated mean method.

\subsection{Median Method} \label{section:median}
\vspace{0.04in}
One final option was to change from the truncated mean to the median of the bins.  In this method, the truncation occurred at both the upper and lower ends of the ordered list of bins (or DOMs, in the case of the DOMs method), such that only one bin (DOM) remained.  The $dE/dx$ value for this bin (DOM) became the value for the event.  For both the binning method and the DOMs method, which had separate optimizations, the energy resolution and RMS were nearly unchanged for all muon energies when compared to the earlier truncated mean methods.   

\section{Results and Conclusions}

\begin{figure}[!h]
\centering
\includegraphics[width=0.48\textwidth]{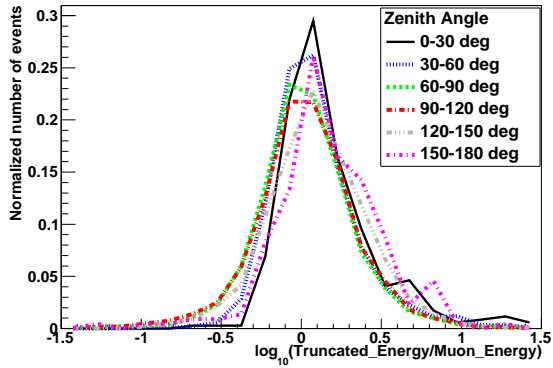} 
\caption{(color online) Energy resolution (in $\log_{10}(E_{\mu})$) for zenith angle bins of 30 degrees for $E_{\mu}$ from 1~TeV to 100~TeV, using the truncated bins method. There is no visible correlation between zenith angle and energy resolution, for the energy range from 1~TeV to 1~EeV using an $E_{\nu}^{-1}$ spectrum. }
\label{fig:zenith}
\end{figure}

\begin{figure}[!h]
\centering
\includegraphics[width=0.48\textwidth]{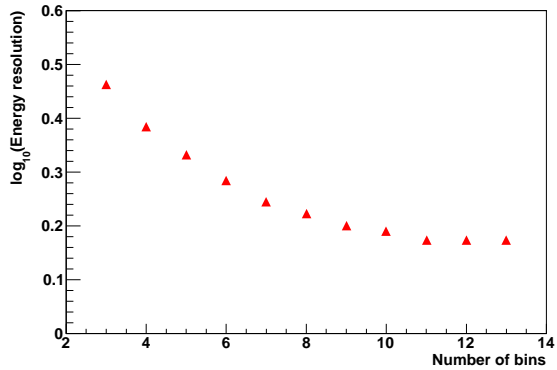} 
\caption{(color online) Correlation between the number of bins and the average energy resolution (in $\log_{10}(E_{\mu})$) of the muon events, using the optimized truncated mean method, for $E_{\mu}$ $\scriptstyle \gtrsim$~10~TeV. }
\label{fig:numbins}
\end{figure}

There is no apparent correlation between the energy calculation and the zenith angle (see Fig.~\ref{fig:zenith}). The distributions of the energy residuals are very similar in shape and width.  There is also no apparent correlation between the energy calculation and azimuth angle, optical properties of the medium, or other parameters.  Thus the truncation method is universally applicable to all particle track zenith and azimuth angles within the detector, with the proper bin size. The energy resolution improves with an increasing number of bins as expected, as shown in Fig.~\ref{fig:numbins}, but levels off at 0.18 in $\log_{10}(E_{\mu})$.   
\begin{figure}[!h]
\centering
\includegraphics[width=0.48\textwidth]{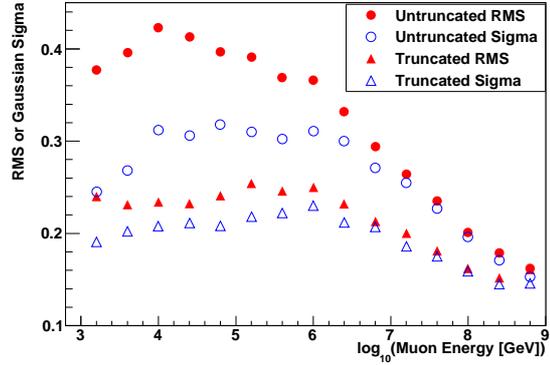} 
\caption{(color online) Comparison of the RMS and Gaussian $\sigma$ of untruncated and truncated $dE/dx$ values vs.~$\log_{10} (E_{\mu}$), for simulated energies between 1~TeV and 1~EeV.  The energy resolution for the truncated method is much improved over the untruncated method.  }
\label{fig:calc_energy_res}
\end{figure}

Figure~\ref{fig:calc_energy_res} compares the RMS and Gaussian $\sigma$ values of truncated and untruncated $dE/dx$ for various muon energies, showing the improvement from using the truncated method.

\begin{figure}[!h]
\centering
\includegraphics[width=0.48\textwidth]{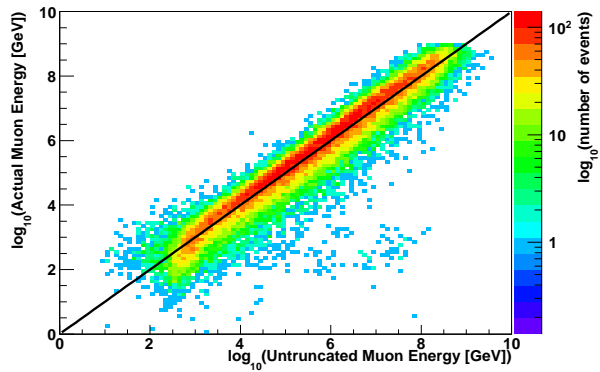} 
\caption{(color online) Actual muon energy versus calculated muon energy for the untruncated method with the 3-equation fit.  The black line is a perfect 1:1 correspondence. }
\label{fig:calc_energy_orig}
\end{figure}

\begin{figure}[!h]
\centering
\includegraphics[width=0.48\textwidth]{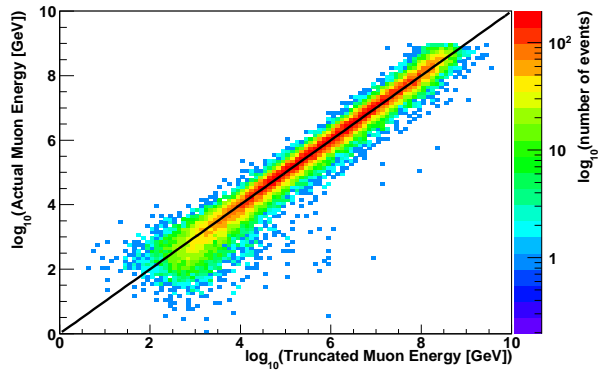} 
\caption{(color online) Actual muon energy versus calculated muon energy for the truncated method (40\% cuts with optimizations).  The black line is a perfect 1:1 correspondence. }
\label{fig:calc_energy_bins}
\end{figure}

Figures~\ref{fig:calc_energy_orig} and \ref{fig:calc_energy_bins} contrast the untruncated and truncated methods, respectively, for actual versus calculated muon energy.  The improvement in energy resolution is fairly uniform over the energy range.  

\begin{figure}[!h]
\centering
\includegraphics[width=0.48\textwidth]{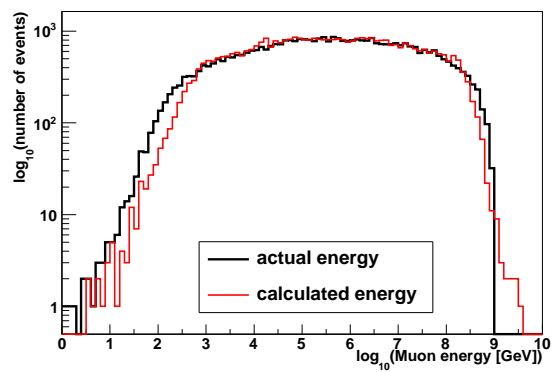} 
\caption{(color online) Comparison of the truncated energy spectrum to the simulated $E_{\nu}^{-1}$ muon energy spectrum.  The agreement is quite good between 1~TeV and 1~EeV, with a slight disagreement at low energy due to the large spread in $dE/dx$ and off-peak fit equation to convert to $E_{\mu}$. }
\label{fig:spectrum}
\end{figure}

Figure~\ref{fig:spectrum} compares the input muon energy spectrum to the spectrum of calculated energies from the truncated method.  The agreement is quite good, with a slight shift in energy below 1~TeV due to the off-peak fit equation plus the spread in $dE/dx$ at low energies.  However, there are no glaring discontinuities in the calculated spectrum, which indicates that the truncated method gives back the original spectrum quite well.  

By using the truncated mean method, the energy resolution is significantly improved. The best truncated method incorporated the following criteria: (1) only include the photoelectrons from optical modules within 10 to 80 m of the track, (2) truncate the highest 40\% of the bins, (3) use both hit and unhit optical modules in the calculation, and (4) sum the remaining photoelectrons separately (observed and expected) to determine the new truncated $dE/dx$ value.  With these optimizations, the energy resolution was improved from 0.29 in $\log_{10}(E_{\mu})$ to 0.22, for the energy range of 1~TeV to 1~EeV.  This is a 26\% improvement in the overall energy resolution, with better resolutions above 10~TeV.  The technique is applicable to any detector that uses $dE/dx$ as the primary means of the particle's energy determination. 


\section{Acknowledgements}

We acknowledge the support from the following agencies: U.S. National Science Foundation-Office of Polar Programs, U.S. National Science Foundation-Physics Division, University of Wisconsin Alumni Research Foundation, the Grid Laboratory Of Wisconsin (GLOW) grid infrastructure at the University of Wisconsin - Madison, the Open Science Grid (OSG) grid infrastructure; U.S. Department of Energy, and National Energy Research Scientific Computing Center, the Louisiana Optical Network Initiative (LONI) grid computing resources, and the National Defense Science and Engineering Graduate (NDSEG) Fellowship Program; National Science and Engineering Research Council of Canada; Swedish Research Council, Swedish Polar Research Secretariat, Swedish National Infrastructure for Computing (SNIC), and Knut and Alice Wallenberg Foundation, Sweden; German Ministry for Education and Research (BMBF), Deutsche Forschungsgemeinschaft (DFG), Research Department of Plasmas with Complex Interactions (Bochum), Germany; Fund for Scientific Research (FNRS-FWO), FWO Odysseus programme, Flanders Institute to encourage scientific and technological research in industry (IWT), Belgian Federal Science Policy Office (Belspo); University of Oxford, United Kingdom; Marsden Fund, New Zealand; Australian Research Council; Japan Society for Promotion of Science (JSPS); the Swiss National Science Foundation (SNSF), Switzerland. 













\end{document}